\documentclass
[twocolumn,amsmath,amssymb,amsfonts,prl,superscriptaddress]{revtex4}%
\usepackage[dvips]{graphicx}
\usepackage{dcolumn}
\usepackage{bm}
\usepackage{amsmath}
\usepackage{amsfonts}
\usepackage{amssymb}
\usepackage{hyperref}
\providecommand{\U}[1]{\protect\rule{.1in}{.1in}}
\begin{document}
\title{Observing Majorana Bound States in \textit{p}-wave Superconductors Using Noise
Measurements in Tunneling Experiments}
\author{C.~J.~Bolech}
\affiliation{Physics Department, Harvard University, Cambridge MA-02138}
\affiliation{Physics\&Astronomy Department, Rice University, Houston TX-77005}
\author{Eugene~Demler}
\affiliation{Physics Department, Harvard University, Cambridge MA-02138}
\date{July 6$^{\text{th}}$, 2006}

\begin{abstract}
The zero-energy bound states at the edges or vortex cores of chiral
\textit{p}-wave superconductors should behave like Majorana fermions. We
introduce a model Hamiltonian that describes the tunnelling process when
electrons are injected into such states. Using a non-equilibrium Green
function formalism, we find exact analytic expressions for the tunnelling
current and noise and identify experimental signatures of the Majorana nature
of the bound states to be found in the shot noise. We discuss the results in
the context of different candidate materials that support triplet
superconductivity. Experimental verification of the Majorana character of
midgap states would have important implications for the prospects of
topological quantum computation.

\end{abstract}

\maketitle

When the individual constituents of a many-body system interact non-trivially
with each other, they can give rise to low-energy states in which the
elementary excitations are very different from the original building blocks.
Examples among electronic condensed-matter systems include the spinons and
holons in Luttinger liquids (realized, \textit{e.g.,} in single-wall carbon
nanotubes) or the Laughlin quasiparticles of fractional quantum Hall systems.
In the context of superconductors, Cooper pairs can be seen as a relatively
simple example of such excitations, but more exotic states are also possible.
In the present work we are interested in the case of \textit{p}-wave chiral
superconductors (\textit{i.e.}~with an order parameter of the type $\hat
{p}_{x}\pm i\hat{p}_{y}$), examples of which would be strontium ruthenate
(\textrm{Sr}$_{2}$\textrm{RuO}$_{4}$) \cite{murakawa2004,xia2006} and,
possibly, a number of organic superconductors like the Bechgaard salts
(\textrm{\{TMTSF\}}$_{2}$\textrm{X}; \textrm{X}$=$\textrm{PF}$_{6}%
$,\textrm{ClO}$_{4}$,...) \cite{chemicalreview}, and even heavy fermions
(\textit{e.g.}~\textrm{UPt}$_{3}$) \cite{tou1996}.

Superconductors with \textit{p}-wave orbital symmetry have spin-triplet
pairing and the order parameter is a tensor in spin space rather than a
scalar. This introduces extra freedom and allows for different types of
superconducting phases, first studied and observed in superfluid $^{3}%
$\textrm{He}. In the so called A-phase, Cooper pairs are in a state dubbed
`equal (or parallel) spin pairing' (ESP); all the examples given above are
candidate systems for ESP. Within weak-coupling BCS theory the up- and
down-spin sectors are then independent from each other and the respective
Bogoliubov-de Gennes (BdG) equations are decoupled. Another aspect of the
A-phase of \textit{p}-wave superconductors is that it can support vortex-core
bound states with a spectrum given by $E_{n}=-\omega_{0}\left(  n+\tilde
{n}\right)  $ with $n\in\mathbb{Z}$, $\tilde{n}=0$ and $\omega_{0}$ a
frequency that depends on the details of the vortex profile \cite{kopnin1991}.
In particular, for $n=0$, one notices that the vortices support `zero modes'.
(This should be contrasted with the case of \textit{s}-wave superconductors
for which the vortex bound-state spectrum has again the same form but this
time $\tilde{n}=1/2$, the other of the only two possibilities consistent with
the $E\rightarrow-E$ symmetry of the BdG equations.) A more detailed
consideration of these midgap bound states in the case of ESP reveals that
they have self-adjoint wavefunctions, naturally described as Majorana fermion
modes, and can also be found as edge states \cite{read2000}. Such Majorana
states constitute one more example of an exotic low-energy collective
excitation that is very different from the original electrons that condensed
into the superconducting state.

It would be already extremely interesting to be able to experimentally
identify these strange Majorana bound states, since that would constitute a
stringent test of our current picture of EPS \textit{p}-wave
superconductivity, but the motivations run further. The availability of
Majorana fermions can be exploited in the context of \emph{quantum
computation}, a completely new and revolutionary approach to computing that
would mix aspects of the digital and the analog computing paradigms by
exploiting the basic laws of quantum mechanics. Majorana operators (call them
$\eta_{i}$) can be taken in pairs to define standard fermionic operators [say,
$c^{\dagger}=\left(  \eta_{r}+i\eta_{l}\right)  /\sqrt{2}$]; each of these
generates a two-dimensional Hilbert space that can be used to define a
quantum-bit (qubit). Because the two Majoranas can be spatially far apart
(\textit{e.g.}~in two different vortices) and are very different from the
usual fermionic quasiparticles around them, the so defined qubit would be
relatively immune to decoherence \cite{ivanov2001}, which would sidestep one
of the crucial problems faced by the development of quantum computing
hardware. Moreover, it turns out that the usual global gauge symmetry of the
fermi fields is reduced to a discrete $\mathbf{Z}_{2}$ symmetry for the
Majoranas at the core of a vortex (and they can be shown to change sign when a
third vortex moves about encircling them \cite{ivanov2001,stone2006}).
Changing the sign of a single Majorana of the pair that defines a qubit
operates the change $c^{\dagger}\rightleftarrows c$, or, in other words, it
acts as a qubit-flip (or q-NOT) gate, and the $\mathbf{Z}_{2}$ symmetry being
discrete leaves no room for errors. This shows that braiding vortices would
perform quantum-logical operations on the information stored in them, an
approach known as topological quantum computation (for a recent review see Ref.~\onlinecite{DasSarma2006b}).

\begin{figure}
[t]
\begin{center}
\includegraphics[
height=1.579in,
width=3.3533in
]%
{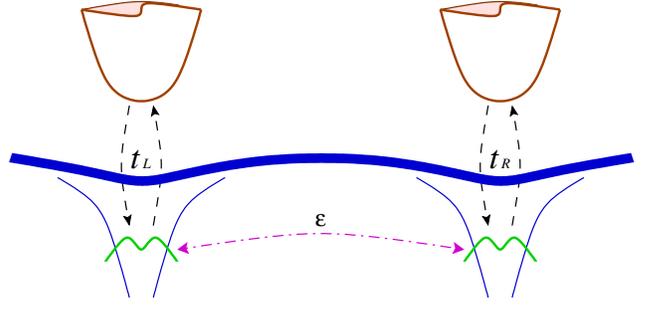}%
\caption{Schematic representation of our model for a setup in which two
contacts (\textit{e.g.} two STM tips) tunnel electrons into the Majorana bound
states at the core of vortices. The two Majorana states might be linked with a
tunnelling amplitude $\varepsilon$. The setup is generic and any or both of
the vortices could be replaced by edge modes or obviated altogether. The model
would also apply to the case of multiband vortices with two different
Majoranas one on each band, in such a case $\varepsilon$ corresponds to the
amplitude of band mixing that can take place at the core of the vortex.}%
\label{Fig:majortun}%
\end{center}
\end{figure}

Recently, the overlap matrix element between a localized electron and a
Majorana bound state was computed for a model of a superconducting wire
(introduced in the context of quantum computation \cite{kitaev2000}) and found
to be non-zero \cite{semenoff2006}. This indicates that tunnelling transport
into Majorana modes is possible, which opens interesting possibilities since
tunnelling has proved repeated times to be an invaluable tool in the study of
superconducting states. The study of tunnelling noise might also be useful,
since shot noise is another probe able to distinguish normal versus
superconducting states and ballistic versus diffusive transport
\cite{fauchere1998,blanter2000}; it can also be sensitive to the charge and
statistics of the carriers and was used, for instance, to confirm the presence
of Laughlin quasiparticles in fractional quantum Hall devices
\cite{saminadayar1997}. Noise probes can be local, in order to study localized
states \cite{birk1995}. The purpose of this article is to model the tunnelling
processes into Majorana bound states and to determine the current and noise
characteristics in order to identify signatures that would allow the
experimental identification and study of such states. A generalized geometry
of the experiment we consider is shown in Fig.~\ref{Fig:majortun}. We shall
find that the Fano factor (or shot noise to current ratio) for such tunnelling
processes has unit matrix structure and is given by%
\begin{equation}
F_{\alpha\beta}\equiv\lim_{V/T\rightarrow\infty}\frac{S_{\alpha\beta}\left(
\omega=0\right)  }{e\left(  I_{\alpha}+I_{\beta}\right)  }=\delta_{\alpha
\beta} \label{eqn:FanoFactor}%
\end{equation}
where $\alpha,\beta=\left\{  L,R\right\}  =\pm1$ label the lead where the
currents ($I_{\alpha,\beta}$) are measured and $S_{\alpha\beta}$ is the noise
spectrum defined below. This is different from the result for a regular
fermionic bound state for which $F_{\alpha\beta}=1/2$ has a `flat' matrix
structure and is half as big --the full expressions for the noise are given in
Eqs.~(\ref{eqn:RLnoise}) and (\ref{eqn:Mnoise})--. We shall argue that measuring the
Fano factor would provide a clear signature of the Majorana nature of a bound state.

The current-voltage characteristics for tunnelling into low-dimensional chiral
\textit{p}-wave superconductors was computed before for the case of planar
junctions using a BTK formalism (including the bound states in a density of
states approximation) \cite{sengupta2001}, or for point contacts using a
microscopic tunnelling Hamiltonian and non-equilibrium Green functions but in
the absence of bound states (\textit{i.e.}~far from vortices or edges)
\cite{bolech2004}. Here we concentrate on the tunneling into bound states or
edge modes for voltages smaller than the superconducting gap and carry out
full microscopic calculations for the current as well as the noise. Our
starting point is the following tunnelling Hamiltonian:%
\[
H=H_{0}+\sqrt{2}t_{L}~i\eta_{l}\left(  \psi_{L}^{\dagger}+\psi_{L}\right)
+\sqrt{2}t_{R}\left(  \psi_{R}^{\dagger}-\psi_{R}\right)  \eta_{r}%
\]
here $\eta_{r,l}$ are two different Majorana operators (located, for instance,
in two different vortices; see Fig.~\ref{Fig:majortun}) and $\psi_{\alpha
}^{\dagger}=\int\psi_{\alpha k}^{\dagger}dk/2\pi$ are the fermions at the
position of the point contact in each of two leads. We consider only the
relevant spin projection of the ESP state and effectively work with spinless
fermions. The $H_{0}$ term in the Hamiltonian has two parts,%
\[
H_{0,\alpha}=\int\frac{dk}{2\pi}~\varepsilon_{k}~\psi_{\alpha k}^{\dagger}%
\psi_{\alpha k}\quad\text{and}\quad H_{0,c}=\varepsilon~i\eta_{l}\eta
_{r}{\small \rightarrow}\varepsilon~c^{\dagger}c~\text{,}%
\]
where $c^{\dagger}$ is defined as above and there is a term $H_{0,\alpha}$ for
each lead. $H_{0,c}$ serves to model the case when the overlap matrix element
between the two Majorana bound states is non-zero
(cf.~Ref.~\onlinecite{semenoff2006}). Notice that the terms in $H$ involving
$\eta$'s are bound to have the form they have due to hermiticity requirements
(cf.~Ref.~\onlinecite{stone2006}). The overlap amplitudes $t_{\alpha}$ have to
be real and we can take them to be positive for the sake of concreteness. The
choice of relative signs in the tunnelling terms is arbitrary and amounts to a
choice of global gauges for the leads.%

For generality, we shall rewrite the Hamiltonian as%
\[
H=H_{0}+\sum_{\alpha=L,R}\left[  t_{\alpha}~\left(  c^{\dagger}\psi_{\alpha
}+\psi_{\alpha}^{\dagger}c\right)  +\alpha\delta_{\alpha}~\left(  \psi
_{\alpha}^{\dagger}c^{\dagger}+c\psi_{\alpha}\right)  \right]  \text{.}%
\]
The Majorana tunnelling case is recovered by setting $\delta_{\alpha
}=t_{\alpha}$, but on the other hand we can set $\delta_{\alpha}=0$ and we
have the standard resonant level model that would describe a regular
(\textit{i.e.}~non-Majorana) edge or bound state \footnote{More precisely, for
Andreev bound states, a toy model for a zero-energy state in a singlet-pairing
superconductor would correspond to $\delta_{\alpha}=t_{\bar{\alpha}}$ and
either $t_{R}$ or $t_{L}$ equal zero. We expect such a model to have similar
shot noise and current-voltage characteristics as a resonant level.}. We can
thus discuss the two cases using a common familiar language and compare them
more easily.

We start by computing the current, which is given by%
\begin{align*}
I\left(  t\right)   &  \equiv\frac{e}{2}\left\langle \partial_{t}\left(
N_{R}-N_{L}\right)  \right\rangle =\frac{e}{2i}\left\langle \left[
H,N_{L}-N_{R}\right]  \right\rangle \\
&  =\frac{e}{4}\sum_{\alpha=L,R}\left[  \alpha t_{\alpha}\left(  G_{\alpha
c}^{K}\mathcal{-}G_{c\alpha}^{K}\right)  +\delta_{\alpha}\left(  F_{\alpha
c}^{K}\mathcal{-}F_{c\alpha}^{\dagger K}\right)  \right]
\end{align*}
where $G_{\alpha c}^{K}$ is the normal Keldysh component of the equal-time
Green function, $iG_{\alpha c}^{K}=\left\langle \psi_{\alpha}c^{\dagger
}\right\rangle _{\text{kel}}$, and $F_{\alpha c}^{K}$ is its anomalous
counterpart, $iF_{\alpha c}^{K}=\left\langle \psi_{\alpha}c\right\rangle
_{\text{kel}}$ (and similarly,\textit{ mutatis mutandis}, for other
components). In order to compute these Green functions we follow the `local
action approach' of Ref.~\onlinecite{bolech2005}, with the main difference
that here the calculation is done fully analytically. The case of
$\delta_{\alpha}=0$ is relatively straight forward and was discussed before in
the literature (see \cite{blanter2000} for a review); we therefore give here
only the results for the Green functions when $\delta_{\alpha}=t_{\alpha}$. In
order to compute them we work with positive frequencies and adopt the spinor
basis given by%
\[
\Psi(\omega)=%
\begin{pmatrix}
\psi_{L}\left(  \omega\right)  & \psi_{L}^{\dagger}\left(  \bar{\omega}\right)
& \psi_{R}\left(  \omega\right)  & \psi_{R}^{\dagger}\left(  \bar{\omega
}\right)  & c\left(  \omega\right)  & c^{\dagger}\left(  \bar{\omega}\right)
\end{pmatrix}
^{T}%
\]
(where the bars stand for minus signs). Inverting the local action --which is
equivalent to solving the full non-equilibrium Dyson equations-- and
restricting ourselves to the symmetric case ($t_{L}=t_{R}=t$), we obtain the
following Green functions: (i) the localized-states Green functions,%
\begin{align*}
G_{cc}^{K}  &  =\frac{1}{4}\left\{  -iD_{c}\left(  \omega-\varepsilon\right)
\left(  \operatorname{th}\frac{\omega-\mu_{\gamma}}{2T}+\operatorname{th}%
\frac{\omega+\mu_{\gamma}}{2T}\right)  \right\} \\
F_{cc}^{\dagger K}  &  =\frac{1}{4}\left\{  -i\gamma F_{c}\left(
\omega\right)  \left(  \operatorname{th}\frac{\omega-\mu_{\gamma}}%
{2T}+\operatorname{th}\frac{\omega+\mu_{\gamma}}{2T}\right)  \right\}
\end{align*}
where we used $\Gamma=\Gamma_{L}+\Gamma_{R}$ with $\Gamma_{\alpha}=\frac{2}%
{W}\left(  t_{\alpha}^{2}+\delta_{\alpha}^{2}\right)  $ (later we will need
also $\breve{\Gamma}=\frac{4}{W}\sum_{\alpha}\alpha\delta_{\alpha}t_{\alpha}$)
to define the functions $D_{c}\left(  \omega\right)  =\frac{2\Gamma}%
{\omega^{2}+\Gamma^{2}}$ and $F_{c}\left(  \omega\right)  =\frac{2\Gamma
}{\omega^{2}-\left(  \varepsilon+i\Gamma\right)  ^{2}}$. The sums on
$\gamma=\left\{  L,R\right\}  $ are implicit and $W=4v_{\text{F}}$
($v_{\text{F}}$ is the fermi velocity in the leads). We introduce now the
notation $\iota_{\beta}=-\iota_{\alpha}=1$ and write, (ii), the inter/intra
lead Green functions,%
\begin{multline*}
G_{\alpha\beta}^{K}=\frac{t^{2}}{W^{2}}\left\{  \delta_{\alpha,\beta}\frac
{32}{i\Gamma}\operatorname{th}\frac{\omega-\mu_{\alpha}}{2T}-i\sum
_{z=\alpha,\beta}\operatorname{th}\frac{\omega-\iota_{z}\mu_{\gamma}}%
{2T}\right. \\
\left.  \times\left[  D_{c}\left(  \omega-\varepsilon\right)  +\beta\gamma
F_{c}^{\ast}\left(  \omega\right)  +\alpha\gamma F_{c}\left(  \omega\right)
+\alpha\beta D_{c}\left(  \omega+\varepsilon\right)  \right]  \right. \\
\left.  -\sum_{z=\alpha,\beta}\left(  \frac{\iota_{z}8}{\omega-\varepsilon
+i\iota_{z}\Gamma}+\frac{\iota_{z}8\alpha\beta}{\omega+\varepsilon+i\iota
_{z}\Gamma}\right)  \operatorname{th}\frac{\omega-\mu_{z}}{2T}\right\}
\end{multline*}%
\begin{multline*}
F_{\alpha\beta}^{\dagger K}=\frac{t^{2}}{W^{2}}\left\{  i\sum_{z=\alpha,\beta
}\operatorname{th}\frac{\omega-\iota_{z}\mu_{\gamma}}{2T}\right. \\
\left.  \times\left[  \alpha D_{c}\left(  \omega-\varepsilon\right)
+\alpha\beta\gamma F_{c}^{\ast}\left(  \omega\right)  +\gamma F_{c}\left(
\omega\right)  +\beta D_{c}\left(  \omega+\varepsilon\right)  \right]  \right.
\\
\left.  +\sum_{z=\alpha,\beta}\left(  \frac{\iota_{z}8\alpha}{\omega
-\varepsilon+i\iota_{z}\Gamma}+\frac{\iota_{z}8\beta}{\omega+\varepsilon
+i\iota_{z}\Gamma}\right)  \operatorname{th}\frac{\omega-\iota_{z}\mu_{z}}%
{2T}\right\}
\end{multline*}
Finally we use the convention that upper/lower indices correspond to
upper/lower signs ($\ast\!\ast$ means there is no complex conjugation) and
write, (iii), the `tunnelling' Green functions,%
\begin{multline*}
G_{%
\genfrac{}{}{0pt}{}{\alpha c}{c\alpha}%
}^{K}=\frac{t}{2W}\left\{  \frac{-8i}{\omega-\varepsilon\mp i\Gamma
}\operatorname{th}\frac{\omega-\mu_{\alpha}}{2T}\right. \\
\left.  \mp\left[  D_{c}\left(  \omega-\varepsilon\right)  +\alpha\gamma
F_{c}^{%
\genfrac{}{}{0pt}{}{\ast\!\ast}{\ast}%
}\left(  \omega\right)  \right]  \sum_{z=\alpha,\beta}\operatorname{th}%
\frac{\omega-\iota_{z}\mu_{\gamma}}{2T}\right\}
\end{multline*}%
\begin{multline*}
F_{%
\genfrac{}{}{0pt}{}{\alpha c}{c\alpha}%
}^{\dagger K}=\frac{t}{2W}\left\{  \frac{\alpha8i}{\pm\omega-\varepsilon
-i\Gamma}\operatorname{th}\frac{\omega\pm\mu_{\alpha}}{2T}\right. \\
\left.  +\left[  \alpha D_{c}\left(  \omega\mp\varepsilon\right)  +\gamma
F_{c}\left(  \omega\right)  \right]  \sum_{z=\alpha,\beta}\operatorname{th}%
\frac{\omega-\iota_{z}\mu_{\gamma}}{2T}\right\}
\end{multline*}

We now use the third set of Green functions and replace in the formula for the
current. Let us first quote the result for a resonant level or double barrier
(cf.~Ref.~\onlinecite{blanter2000}),%
\begin{align*}
\frac{I}{e} &  =\frac{\Gamma_{L}\Gamma_{R}}{\Gamma}\sum_{\alpha=R,L}\alpha
\int_{-\infty}^{+\infty}D_{c}\left(  \omega-\varepsilon\right)
\operatorname{th}\frac{\omega-\mu_{\alpha}}{2T}\frac{d\omega}{2\pi}\\
&  \underset{T\rightarrow0}{\longrightarrow}\frac{2}{\pi}\frac{\Gamma
_{L}\Gamma_{R}}{\Gamma}\sum_{\alpha=R,L}\alpha\arctan\frac{\varepsilon
-\mu_{\alpha}}{\Gamma}~\text{.}%
\end{align*}
Notice the current becomes zero if either $\Gamma_{L}$ or $\Gamma_{R}$ vanish;
in fact, we always find $I_{L}=I_{R}$ (with $I_{\alpha}\equiv e\alpha
\left\langle \partial_{t}N_{\alpha}\right\rangle $). The situation is
different when the tunnelling is into a Majorana state; in such a case, the
two currents are \textit{independent} (parametrically related if
$\varepsilon\neq0$) and given by%
\[
\frac{I_{\alpha}}{e}=\alpha\Gamma_{\alpha}\int_{-\infty}^{+\infty}D_{c\alpha
}\left(  \omega\right)  \operatorname{th}\frac{\omega-\mu_{\alpha}}{2T}%
\frac{d\omega}{2\pi}%
\]
with (valid also if $t_{L}\neq t_{R}$)%
\[
D_{c\alpha}\left(  \omega\right)  =\frac{4\omega^{2}\Gamma-2\left(
\Gamma-\alpha\breve{\Gamma}\right)  \left(  \omega^{2}-\varepsilon^{2}%
-\Gamma^{2}+\breve{\Gamma}^{2}\right)  }{\left(  \omega^{2}-\varepsilon
^{2}-\Gamma^{2}+\breve{\Gamma}^{2}\right)  ^{2}+4\Gamma^{2}\omega^{2}%
}~\text{.}%
\]
If $\Gamma_{L}=\Gamma_{R}$ and $\varepsilon=0$, then $I=\left(  I_{L}%
+I_{R}\right)  /2$ coincides with the result for a resonant level. Even though
the results are mathematically different (notice that for the case of a wire,
$\varepsilon\neq0$ always regardless of its size \cite{semenoff2006}), the
differences might be hard to detect experimentally; but we shall now see that
the noise provides a more robust signature of the Majorana nature of the
tunnelling intermediate states.

The noise is a measure of the deviations of the current from its average value
[$\Delta I_{\alpha}\left(  t\right)  =I_{\alpha}\left(  t\right)
-\left\langle I_{\alpha}\left(  t\right)  \right\rangle $] and it is standard
to define it as the following \emph{correlator }\cite{blanter2000}:%
\[
S_{\alpha\beta}\left(  t-t^{\prime}\right)  =\frac{1}{2}\left\langle \Delta
I_{\alpha}\left(  t\right)  \Delta I_{\beta}\left(  t^{\prime}\right)  +\Delta
I_{\beta}\left(  t^{\prime}\right)  \Delta I_{\alpha}\left(  t\right)
\right\rangle ~\text{.}%
\]
Its Fourier transform is known as the noise power spectrum and for
$\delta_{\alpha}=0$ is given by%
\begin{multline*}
S_{\alpha\beta}\left(  \omega\right)  =-\frac{\alpha\beta t_{\alpha}t_{\beta}%
}{2}\left[  G_{\alpha c}^{K}\circ G_{\beta c}^{K}+G_{c\beta}^{K}\circ
G_{c\alpha}^{K}\right.  \\
\left.  -G_{\alpha\beta}^{K}\circ G_{cc}^{K}-G_{cc}^{K}\circ G_{\beta\alpha
}^{K}\right]
\end{multline*}
For the general case, the expression is similar but longer; it involves also
anomalous Green functions and comprises thirty two terms with a mix of both
correlation \footnote{$\left[  G_{1}\circ G_{2}\right]  \left(  \omega\right)
\equiv\int\frac{d\omega^{\prime}}{2\pi}G_{1}\left(  \omega^{\prime}%
+\frac{\omega}{2}\right)  G_{2}\left(  \omega^{\prime}-\frac{\omega}%
{2}\right)  $.} and convolution products.

We concentrate on the zero-frequency noise component and first rederive the
result for a resonant level \cite{chen1991,averin1993}:%
\begin{multline*}
\frac{S_{\alpha\beta}\left(  \omega=0\right)  }{e^{2}}=\operatorname{cth}%
\left(  \frac{eV}{2T}\right)  \left\{  \left(  2-\frac{4\Gamma_{R}\Gamma_{L}%
}{\Gamma^{2}}\right)  \frac{I}{e}\right. \\
\left.  -\frac{\Gamma^{2}}{2\pi}\left(  \frac{4\Gamma_{R}\Gamma_{L}}%
{\Gamma^{2}}\right)  ^{2}\left[  \frac{\omega^{\prime}-\varepsilon}{\left(
\omega^{\prime}-\varepsilon\right)  ^{2}+\Gamma^{2}}\right]  _{\mu_{R}}%
^{\mu_{L}}\right\}
\end{multline*}
where $eV=\mu_{L}-\mu_{R}$. In the fully-symmetric case ($\mu_{L}=-\mu_{R}%
=\mu$ and $t_{L}=t_{R}=t$) it takes the form%
\begin{equation}
\frac{S_{\alpha\beta}\left(  0\right)  }{e^{2}}\rightarrow\operatorname{cth}%
\left(  \frac{eV}{2T}\right)  \left\{  \frac{I}{e}-\frac{\Gamma^{2}}{2\pi
}\left[  \frac{\omega^{\prime}-\varepsilon}{\left(  \omega^{\prime
}-\varepsilon\right)  ^{2}+\Gamma^{2}}\right]  _{-\mu}^{+\mu}\right\}
\label{eqn:RLnoise}%
\end{equation}
and the Fano factor becomes $F_{\alpha\beta}=1-2\Gamma_{R}\Gamma_{L}%
/\Gamma^{2}\rightarrow1/2$ as mentioned earlier. On the other hand, when the
tunnelling takes place into a Majorana bound state we find the following
result (for the fully-symmetric case):%
\begin{multline*}
\frac{S_{\alpha\beta}\left(  \omega=0\right)  }{e^{2}}=\operatorname{cth}%
\left(  \frac{eV}{2T}\right)  \left\{  2\delta_{\alpha\beta}\frac{I}{e}\right.
\\
\left.  -\frac{\Gamma^{2}}{2\pi}\left[  \frac{\left(  \omega^{\prime
}-\varepsilon\right)  }{\left(  \omega^{\prime}-\varepsilon\right)
^{2}+\Gamma^{2}}\right]  _{-\mu}^{+\mu}-\alpha\beta\frac{\Gamma^{2}}%
{4\pi\varepsilon}\ln\frac{\left(  \varepsilon+\mu\right)  ^{2}+\Gamma^{2}%
}{\left(  \varepsilon-\mu\right)  ^{2}+\Gamma^{2}}\right\}
\end{multline*}
Notice that the diagonal and off-diagonal matrix components of $S_{\alpha
\beta}$ are different now. In particular, we remark that $\lim_{\varepsilon
\rightarrow0}S_{\alpha\bar{\alpha}}=0$. Taken together with the result given
above for the current, this indicates that in the $\varepsilon\rightarrow0$
limit the right and left tunnelling processes are completely independent even
at the level of current fluctuations. It is therefore instructive and
important to study this case more in detail, because of its greater simplicity
and its relevance to single-tip setups. We relax the condition on the chemical
potentials (\textit{i.e.}~consider $\mu_{L},\mu_{R}$ arbitrary) and find that
the noise can be written as%
\begin{equation}
\frac{S_{\alpha\beta}\left(  0\right)  }{e^{2}}\rightarrow\delta_{\alpha\beta
}\operatorname{cth}\left(  \frac{\mu_{\alpha}}{T}\right)  \left\{
2\frac{I_{\alpha}}{e}-\frac{4\Gamma_{\alpha}^{2}}{\pi}\frac{2\mu_{\alpha}}%
{\mu_{\alpha}^{2}+4\Gamma_{\alpha}^{2}}\right\}  \text{.} \label{eqn:Mnoise}%
\end{equation}
Given the right-left independence, we expect the expression to be valid also
when $t_{L}\neq t_{R}$. In particular, this implies that the Fano factor,
Eq.~(\ref{eqn:FanoFactor}), is not sensitive to the contacts asymmetry, unlike
what happens for $\delta_{R,L}=0$.

Presently, the likely best place to find single-Majorana bound states is at
the edges of Bechgaard salts. In the presence of magnetic fields much larger
than the paramagnetic limit, but still smaller than $H_{c2}$, the
superconducting state ought to be similar to the A1-phase of $^{3}$\textrm{He}
in which all the spins are aligned with the magnetic field and effectively
there is only one spin species. The situation in the case of strontium
ruthenate is more complex and the existence of single-Majorana bound states
seems more elusive. Let us start by pointing out that the vortices that
support isolated Majorana zero modes are no less bizarre themselves: they
carry only half of the superconducting flux quantum and the vorticity lies
entirely in one of the spin sectors while the other one does not show a
winding phase at all \cite{DasSarma2006a}. Recent scanning tunnelling
microscope (STM) studies of \textrm{Sr}$_{2}$\textrm{RuO}$_{4}$ have found a
square lattice of vortices with a full flux quantum each \cite{lupien2006},
even though the magnetic field was presumably large enough to take the system
into an ESP state \cite{murakawa2004,DasSarma2006b}. However, the experiments
did find a strong zero bias conductance peak that remains unexplained. One
possibility is that the observed vortices are built out of two half-vortices,
one for each spin projection. In that case, one would expect, respectively,
two Majorana bound states, and $\varepsilon$ in our model would be related to
the amplitude of spin mixing at the vortex core. If this is the case, the
vortices would not have the same braiding properties as the half-vortices, but
the Fano factor would nevertheless be sensitive to the Majorana nature of the
midgap states. Whether these `double half-vortices' would still be useful for
topological quantum computation would require further investigation, but
identifying them experimentally would be extremely interesting in any case.

\begin{acknowledgments}
We would like to acknowledge discussions with E.~Altman, B.~Halperin,
J.~Hoffmann, A.~Kolezhuk and A.~Polkovnikov. This work was partially supported
by the NSF (DMR-0132874).
\end{acknowledgments}


\begin{thebibliography}{99}
\expandafter\ifx\csname natexlab\endcsname\relax\def\natexlab#1{#1}\fi
\expandafter\ifx\csname bibnamefont\endcsname\relax
  \def\bibnamefont#1{#1}\fi
\expandafter\ifx\csname bibfnamefont\endcsname\relax
  \def\bibfnamefont#1{#1}\fi
\expandafter\ifx\csname citenamefont\endcsname\relax
  \def\citenamefont#1{#1}\fi
\expandafter\ifx\csname url\endcsname\relax
  \def\url#1{\texttt{#1}}\fi
\expandafter\ifx\csname urlprefix\endcsname\relax\def\urlprefix{URL }\fi
\providecommand{\bibinfo}[2]{#2}
\providecommand{\eprint}[2][]{\url{#2}}

\bibitem[Murakawa et~al.(2004)Murakawa, Ishida, Kitagawa, Mao, and
Maeno]{murakawa2004}\bibinfo{author}{\bibfnamefont{H.}~\bibnamefont{Murakawa}}
\textit{et al.}, \bibinfo{journal}{Phys. Rev. Lett.}
\textbf{\bibinfo{volume}{97}}, \bibinfo{pages}{167002} (\bibinfo{year}{2006}).

\bibitem[Xia et~al.(2006)Xia, Maeno, Beyersdorf, Fejer, and Kapitulnik]%
{xia2006}\bibinfo{author}{\bibfnamefont{J.}~\bibnamefont{Xia}} \textit{et al.},
\bibinfo{journal}{Phys. Rev. Lett.} \textbf{\bibinfo{volume}{97}},
\bibinfo{pages}{167002} (\bibinfo{year}{2006}).

\bibitem[che(2004)]{chemicalreview}\bibinfo{journal}{Chem. Rev.}
\textbf{\bibinfo{volume}{104}} (\bibinfo{year}{2004}),
\bibinfo{note}{issue on Molecular
Conductors}.

\bibitem[Tou et~al.(1996)Tou, Kitaoka, Asayama, Kimura,
\ifmmode~\={O}\else\={O}\fi{}nuki, Yamamoto, and Maezawa]{tou1996}%
\bibinfo{author}{\bibfnamefont{H.}~\bibnamefont{Tou}} \textit{et al.},
\bibinfo{journal}{Phys. Rev. Lett.} \textbf{\bibinfo{volume}{77}},
\bibinfo{pages}{1374} (\bibinfo{year}{1996}).

\bibitem[Kopnin and Salomaa(1991)]{kopnin1991}%
\bibinfo{author}{\bibfnamefont{N.~B.} \bibnamefont{Kopnin}} and
\bibinfo{author}{\bibfnamefont{M.~M.} \bibnamefont{Salomaa}},
\bibinfo{journal}{Phys. Rev. B} \textbf{\bibinfo{volume}{44}},
\bibinfo{pages}{9667} (\bibinfo{year}{1991}).

\bibitem[Read and Green(2000)]{read2000}%
\bibinfo{author}{\bibfnamefont{N.}~\bibnamefont{Read}} and
\bibinfo{author}{\bibfnamefont{D.}~\bibnamefont{Green}},
\bibinfo{journal}{Phys. Rev. B} \textbf{\bibinfo{volume}{61}},
\bibinfo{pages}{10267} (\bibinfo{year}{2000}).

\bibitem[Ivanov(2001)]{ivanov2001}%
\bibinfo{author}{\bibfnamefont{D.~A.} \bibnamefont{Ivanov}},
\bibinfo{journal}{Phys. Rev. Lett.} \textbf{\bibinfo{volume}{86}},
\bibinfo{pages}{268} (\bibinfo{year}{2001}).

\bibitem[Stone and {S.-B.~Chung}(2006)]{stone2006}%
\bibinfo{author}{\bibfnamefont{M.}~\bibnamefont{Stone}} and
\bibinfo{author}{\bibnamefont{{S.-B.~Chung}}}, \bibinfo{journal}{Phys. Rev.
B} \textbf{\bibinfo{volume}{73}}, \bibinfo{pages}{014505} (\bibinfo{year}{2006}).

\bibitem[{Das Sarma} et~al.(2006){Das Sarma}, Freedman, and Nayak]%
{DasSarma2006b}\bibinfo{author}{\bibfnamefont{S.}~\bibnamefont{{Das Sarma}}},
\bibinfo{author}{\bibfnamefont{M.}~\bibnamefont{Freedman}}, and
\bibinfo{author}{\bibfnamefont{C.}~\bibnamefont{Nayak}},
\bibinfo{journal}{Phys. Today} \textbf{\bibinfo{volume}{59}}, No.~7,
\bibinfo{pages}{32} (\bibinfo{year}{2006}).

\bibitem[{A.~Yu. Kitaev}(2000)]{kitaev2000}%
\bibinfo{author}{\bibnamefont{{A.~Yu. Kitaev}}}, \bibinfo{note}{arXiv:cond-mat/0010440}.

\bibitem[Semenoff and Sodano(2006)]{semenoff2006}%
\bibinfo{author}{\bibfnamefont{G.~W.} \bibnamefont{Semenoff}} and
\bibinfo{author}{\bibfnamefont{P.}~\bibnamefont{Sodano}}, \bibinfo{note}{arXiv:cond-mat/0601261}.

\bibitem[Fauch\`ere et~al.(1998)Fauch\`ere, Lesovik, and Blatter]%
{fauchere1998}\bibinfo{author}{\bibfnamefont{A.~L.} \bibnamefont{Fauch\`ere}},
\bibinfo{author}{\bibfnamefont{G.~B.} \bibnamefont{Lesovik}}, and
\bibinfo{author}{\bibfnamefont{G.}~\bibnamefont{Blatter}},
\bibinfo{journal}{Phys. Rev. B} \textbf{\bibinfo{volume}{58}},
\bibinfo{pages}{11177} (\bibinfo{year}{1998}).

\bibitem[Blanter and B{\"{u}}ttiker(2000)]{blanter2000}%
\bibinfo{author}{\bibfnamefont{Y.~M.} \bibnamefont{Blanter}} and
\bibinfo{author}{\bibfnamefont{M.}~\bibnamefont{B{\"u}ttiker}},
\bibinfo{journal}{Phys. Rep.} \textbf{\bibinfo{volume}{336}},
\bibinfo{pages}{1} (\bibinfo{year}{2000}).

\bibitem[Saminadayar et~al.(1997)Saminadayar, Glattli, Jin, and Etienne]%
{saminadayar1997}\bibinfo{author}{\bibfnamefont{L.}~\bibnamefont{Saminadayar}}
\textit{et al.}, \bibinfo{journal}{Phys. Rev. Lett.}
\textbf{\bibinfo{volume}{79}}, \bibinfo{pages}{2526} (\bibinfo{year}{1997}).

\bibitem[Birk et~al.(1995)Birk, de~Jong, and Sch\"onenberger]{birk1995}%
\bibinfo{author}{\bibfnamefont{H.}~\bibnamefont{Birk}},
\bibinfo{author}{\bibfnamefont{M.~J.~M.} \bibnamefont{de~Jong}}, and
\bibinfo{author}{\bibfnamefont{C.}~\bibnamefont{Sch\"onenberger}},
\bibinfo{journal}{Phys. Rev. Lett.} \textbf{\bibinfo{volume}{75}},
\bibinfo{pages}{1610} (\bibinfo{year}{1995}).

\bibitem[Sengupta et~al.(2001)Sengupta, {\v{Z}}uti\'{c}, Kwon, Yakovenko, and
{Das Sarma}]{sengupta2001}%
\bibinfo{author}{\bibfnamefont{K.}~\bibnamefont{Sengupta}} \textit{et al.},
\bibinfo{journal}{Phys. Rev. B} \textbf{\bibinfo{volume}{63}},
\bibinfo{pages}{144531} (\bibinfo{year}{2001}).

\bibitem[Bolech and Giamarchi(2004)]{bolech2004}%
\bibinfo{author}{\bibfnamefont{C.~J.} \bibnamefont{Bolech}} and
\bibinfo{author}{\bibfnamefont{T.}~\bibnamefont{Giamarchi}},
\bibinfo{journal}{Phys. Rev. Lett.} \textbf{\bibinfo{volume}{92}},
\bibinfo{pages}{127001} (\bibinfo{year}{2004}).

\bibitem[Bolech and Giamarchi(2005)]{bolech2005}%
\bibinfo{author}{\bibfnamefont{C.~J.} \bibnamefont{Bolech}} and
\bibinfo{author}{\bibfnamefont{T.}~\bibnamefont{Giamarchi}},
\bibinfo{journal}{Phys. Rev. B} \textbf{\bibinfo{volume}{71}},
\bibinfo{pages}{024517} (\bibinfo{year}{2005}).

\bibitem[Chen and Ting(1991)]{chen1991}%
\bibinfo{author}{\bibfnamefont{L.~Y.} \bibnamefont{Chen}} and
\bibinfo{author}{\bibfnamefont{C.~S.} \bibnamefont{Ting}},
\bibinfo{journal}{Phys. Rev. B} \textbf{\bibinfo{volume}{43}},
\bibinfo{pages}{4534} (\bibinfo{year}{1991}).

\bibitem[Averin(1993)]{averin1993}%
\bibinfo{author}{\bibfnamefont{D.~V.} \bibnamefont{Averin}},
\bibinfo{journal}{J. Applied Phys.} \textbf{\bibinfo{volume}{73}},
\bibinfo{pages}{2593} (\bibinfo{year}{1993}).

\bibitem[Sarma et~al.(2006)Sarma, Nayak, and Tewari]{DasSarma2006a}%
\bibinfo{author}{\bibfnamefont{S.} \bibnamefont{Das Sarma}},
\bibinfo{author}{\bibfnamefont{C.}~\bibnamefont{Nayak}}, and
\bibinfo{author}{\bibfnamefont{S.}~\bibnamefont{Tewari}},
\bibinfo{journal}{Phys. Rev. B} \textbf{\bibinfo{volume}{73}},
\bibinfo{pages}{220502(R)} (\bibinfo{year}{2006}).

\bibitem[Lupien et~al.(2005)Lupien, Dutta, Barker, Maeno, and Davis]%
{lupien2006}\bibinfo{author}{\bibfnamefont{C.}~\bibnamefont{Lupien}}
\textit{et al.}, \bibinfo{note}{arXiv:cond-mat/0503317}.
\end{thebibliography}

\end{document}